# Generalisation of Helmholtz-Thèvenin theorem to three-phase electrical circuits

**Gheorghe MIHAI**

*University of Craiova, Faculty of Electrotechnics, Ddul Decebal 5, Romania*
*E-mail:* gmihai@elth.ucv.ro

**Abstract.** The scope of this paper is to determine the generalized form for equivalent tension generator theorem (Helmholtz-Thèvenin theorem) for three-phase electrical circuit. Any complicated electrical power systems we can reduce depending on any three-phase electrical consumer to a three-phase electrical generator that has certain internal impedance. Starting with this assumption, we have demonstrated the way to obtain the electromotive voltages for an equivalent generator and its internal impedances.

## 1. Introduction

From specialized literature is show that a linear and active electrical system, so active linear dipole, can reduce to an electrical generator. Its electromotive voltage is equal with idle running system voltage reporting to **a-b** terminals and its internal impedance equal with measured impedance starting at **a-b** terminals, when the network is passive.

We are starting with following notations, from figure 1.

$\underline{U}_{ab0}$ > The idle running voltage for the linear active dipole reporting to **a-b** terminals.

$\underline{Z}_{abp}$ > The linear impedance of active dipole, looking to **a-b** terminals, then its passive.

$\underline{Z}$ > The charge impedance.

We propose to generalize the Helmholtz-Thèvenin's theorem for a linear active electrical three-phase system, without magnetic couplings, as we see in figure 2.

So, we will demonstrate that we can substitute all three-phase electrical system, looking to terminals consumer, at $\underline{Z}_1$, $\underline{Z}_2$, $\underline{Z}_3$ impedance, with three-phase electrical generator, its electromotive voltages are $\underline{U}_{e1}$, $\underline{U}_{e2}$, $\underline{U}_{e3}$ and its internal phase –impedance are $\underline{Z}_{i1}$, $\underline{Z}_{i2}$, $\underline{Z}_{i3}$, as we see in figure 2.

In these assumptions, we will determinate the parameters of electrical generator in terms of the system' parameters and consumer position.

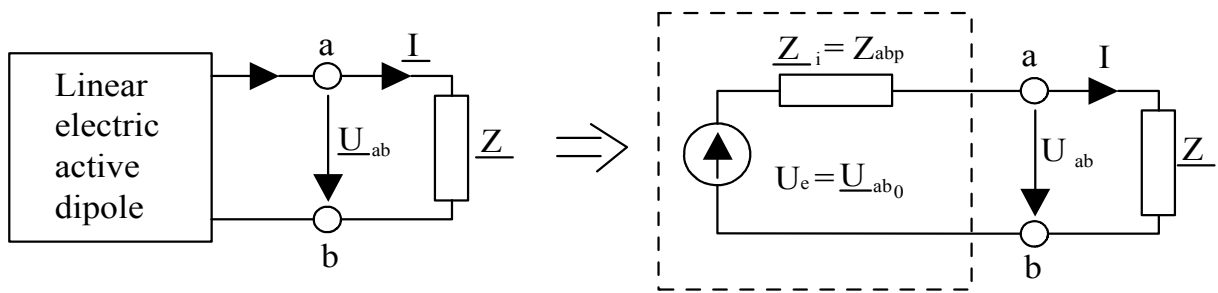

*Figure 1. The Helmholtz-Thèvenin's theorem for linear active electrical dipole*

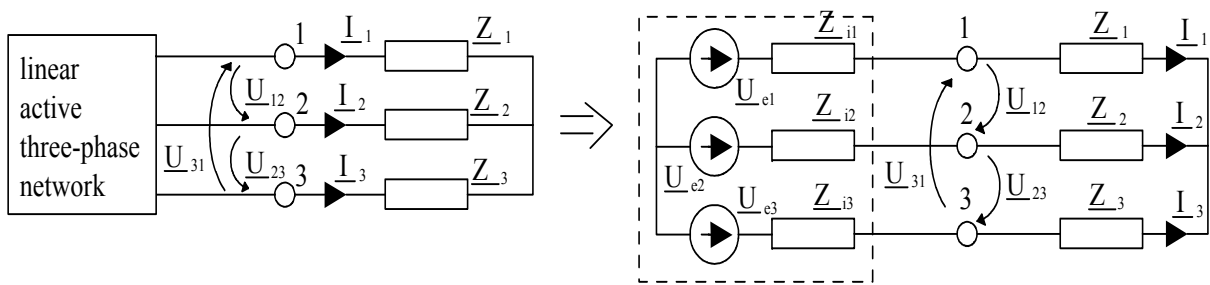

*Figure 2. The generalized Helmholtz-Thevenin's theorem for three-phase electrical circuit.*
*a) an linear three-phase electrical system connected to three-phase consumer*
*b) an equivalent three-phase electrical generator with the system connected to the same three-phase consumer.*

## 2. The parameters determination for equivalent electrical three-phase generator without null wire

By hypothesis, the electrical three-phase system is linear so, between the terminal voltages and the electrical current of consumer is the relation:

$$\underline{\hat{U}} = \underline{\hat{\hat{C}}}_1 \underline{\hat{I}} + \underline{\hat{C}}_2 \qquad (1)$$

Where:

$$\underline{\hat{U}} = \begin{pmatrix} \underline{U}_{12} \\ \underline{U}_{23} \\ \underline{U}_{31} \end{pmatrix} \qquad \underline{\hat{I}} = \begin{pmatrix} \underline{I}_1 \\ \underline{I}_2 \\ \underline{I}_3 \end{pmatrix} \qquad \underline{\hat{\hat{C}}}_1 = \begin{pmatrix} C'_{11} & C'_{12} & C'_{13} \\ C'_{21} & C'_{22} & C'_{23} \\ C'_{31} & C'_{32} & C'_{33} \end{pmatrix} \qquad \underline{\hat{C}}_2 = \begin{pmatrix} C_1^2 \\ C_2^2 \\ C_3^2 \end{pmatrix}$$

In order to determine the constant $\underline{\hat{C}}_2$, we will consider the consumer power off, so that the system idles functioning in relation to the consumer:

From relation (1) we obtain:

$$\underline{\hat{C}}_2 = \begin{pmatrix} \underline{U}_{120} \\ \underline{U}_{230} \\ \underline{U}_{310} \end{pmatrix} \qquad (2)$$

We can also obtain the current values at consumer equal to zero by electrical way, if the consumer is connected to a linear three-phase auxiliary generator, as we see in figure 3.

The phase currents to consumer are the expressions:

$$\begin{cases} \underline{I}_{1r} = \underline{I}_1 + \underline{I'}_1 = 0 \\ \underline{I}_{2r} = \underline{I}_2 + \underline{I'}_2 = 0 \\ \underline{I}_{3r} = \underline{I}_3 + \underline{I'}_3 = 0 \end{cases} \qquad (3)$$

Conforming the superposition principle $\underline{I}_{ir}$ currents from system (3) are obtain from $\underline{I}_1, \underline{I}_2$ and $\underline{I}_3$ currents do to the system if the auxiliary electrical generator is passive and from $\underline{I'}_1$, $\underline{I'}_2, \underline{I'}_3$ electrical currents if the system is passive and working only the auxiliary electrical generator, figure 4.

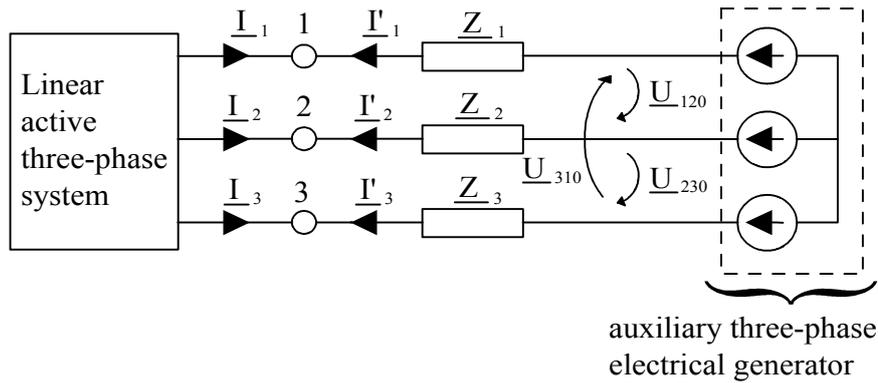

*Figure 3. The electrical connection between the consumer and electrical generator*

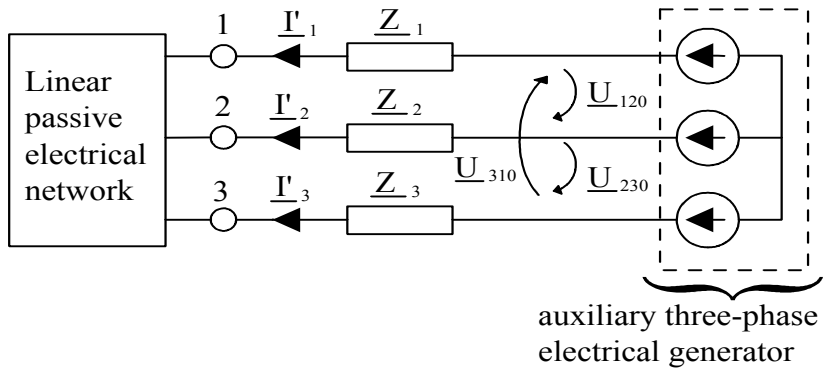

*Figure 4. The electrical schema of a passive system and an auxiliary generator that working*

We are considered the passive electrical system can be reduce to three impedance in Y connected, figure 5.

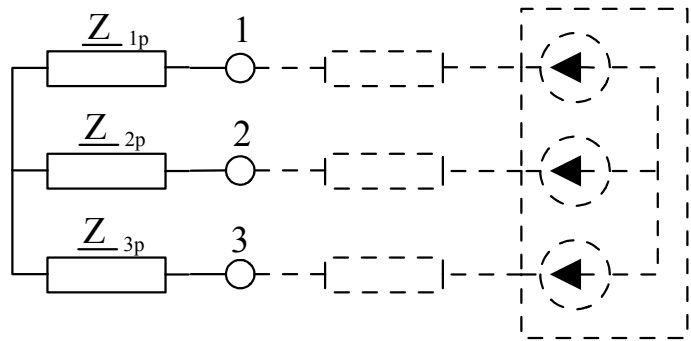

*Figure5. The electrical schema of a passive system reduces to three impedance's Y connected.*

As we show in figure 6, where are three schema of measuring for the impedance's, in relation with 1, 2 and 3 terminals:

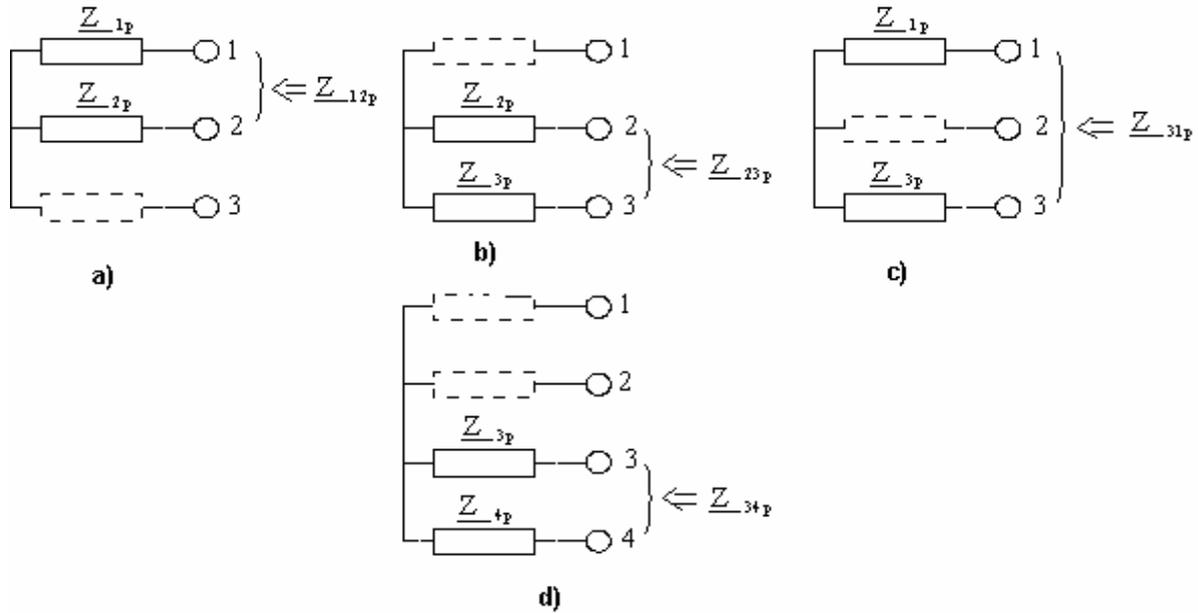

*Figure 6. The schema of measuring impedance in relation with 12, 23, 31 and 34 terminals for Y connection.*

- in relation with 12 terminals, fig. 6a: $\underline{Z}_{12p} = \underline{Z}_{1p} + \underline{Z}_{2p}$

- in relation with 23 terminals, fig.6b: $\underline{Z}_{23p} = \underline{Z}_{2p} + \underline{Z}_{3p}$

- in relation with 31 terminals, fig.6c: $\underline{Z}_{31p} = \underline{Z}_{3p} + \underline{Z}_{1p}$

- in relation with 34 terminals, fig.6d: $\underline{Z}_{34p} = \underline{Z}_{3p} + \underline{Z}_{4p}$

From the last relations we obtain:

$$\begin{cases} \underline{Z}_{1p} = \dfrac{\underline{Z}_{12p} - \underline{Z}_{23p} + \underline{Z}_{31p}}{2} \\[2mm] \underline{Z}_{2p} = \dfrac{\underline{Z}_{23p} - \underline{Z}_{31p} + \underline{Z}_{12p}}{2} \\[2mm] \underline{Z}_{3p} = \dfrac{\underline{Z}_{31p} - \underline{Z}_{12p} + \underline{Z}_{23p}}{2} \\[2mm] \underline{Z}_{4p} = \underline{Z}_{34p} - \underline{Z}_{3p} \end{cases} \qquad (4)$$

From fig.5, we can observe that $\underline{Z}_{1p}, \underline{Z}_{2p}$ and $\underline{Z}_{3p}$ can be consider internal impedance's for auxiliary electrical generator, so:

$$\underline{Z}_{i1} = \underline{Z}_{ip}, \ \ \underline{Z}_{i2} = \underline{Z}_{2p}, \ \underline{Z}_{i3} = \underline{Z}_{3p} \ \text{ and } \ \ \underline{Z}_{i4} = \underline{Z}_{4p} \qquad (5)$$

The equivalent electrical generator has:
- the linear electromotive voltage equal to the linear voltage of the idle running system in the considered point:

$$\begin{cases} \underline{U}_{e12} = \underline{U}_{120} \\ \underline{U}_{e23} = \underline{U}_{230} \\ \underline{U}_{e31} = \underline{U}_{310} \end{cases} \tag{6}$$

- the internal phase impedance equal to the passive phase electrical system, see in the considered point, relation(5).

Returning in relation (1), we can observe that $\hat{\underline{\hat{C}}}_1$ can be obtained from mathematical condition $\hat{\underline{U}} = \mathbf{0}$. This is similar one short-circuit at terminals consumer:

$$\hat{\underline{0}} = \hat{\underline{\hat{C}}}_1 \hat{\underline{I}}_1 + \hat{\underline{C}}_2 \tag{7}$$

For the schema presented in fig.2b and fig.3 and based on the superposition principle, the network has an equivalent electrical scheme:

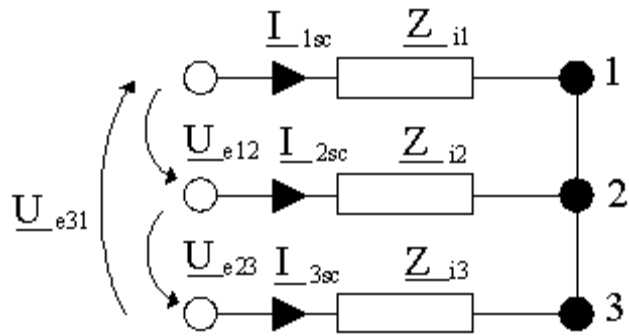

*Figure 7. One short-circuits three-phase to the terminals electrical system*

From the Kirchhoff' second theorem we are obtained:

$$\begin{pmatrix} \underline{U}_{e12} \\ \underline{U}_{e23} \\ \underline{U}_{e31} \end{pmatrix} = \begin{pmatrix} \underline{Z}_{i1} & -\underline{Z}_{i2} & 0 \\ 0 & \underline{Z}_{i2} & -\underline{Z}_{i3} \\ -\underline{Z}_{i1} & 0 & \underline{Z}i3 \end{pmatrix} \cdot \begin{pmatrix} \underline{I}_{1sc} \\ \underline{I}_{2sc} \\ \underline{I}_{3sc} \end{pmatrix} \tag{8}$$

From (7) and (8) relations we are obtained:

$$\hat{\underline{\hat{C}}}_1 = -\begin{pmatrix} \underline{Z}_{i1} & -\underline{Z}_{i2} & 0 \\ 0 & \underline{Z}_{i2} & -\underline{Z}_{i3} \\ -\underline{Z}_{i1} & 0 & \underline{Z}_{i3} \end{pmatrix} = \hat{\underline{\hat{Z}}}_i \tag{9}$$

That is the internal impedance matrix.

## 3. The generalization of Helmholtz-Thèvenin formulae for three-phase electrical systems without null wire

From literature we are the linear active dipole formulae:

$$\underline{U}_{ab0} = (\underline{Z} + \underline{Z}_{abp})\underline{I} \tag{10}$$

Starting with (1), (2), (4), (5) and (9) formula, we are obtained the consumer electrical current relation to an electrical three-phase network:

$$\begin{pmatrix} \underline{U}_{12} \\ \underline{U}_{23} \\ \underline{U}_{31} \end{pmatrix} = \begin{pmatrix} \underline{Z}_1 & -\underline{Z}_2 & \mathbf{0} \\ \mathbf{0} & \underline{Z}_2 & -\underline{Z}_3 \\ -\underline{Z}_1 & \mathbf{0} & \underline{Z}_3 \end{pmatrix} \begin{pmatrix} \underline{I}_1 \\ \underline{I}_2 \\ \underline{I}_3 \end{pmatrix} \qquad (11)$$

In matrices form:

$$\hat{\underline{U}} = \hat{\underline{\underline{Z}}} \hat{\underline{I}} \qquad (12)$$

So, to three-phase electrical circuit, (10) formulae become:

$$\hat{\underline{U}}_0 = (\hat{\underline{\underline{Z}}} + \hat{\underline{\underline{Z}}}_p) \hat{\underline{I}} \qquad (13)$$

Where:

$$\hat{\underline{\underline{Z}}} = \begin{pmatrix} \underline{Z}_1 & -\underline{Z}_2 & \mathbf{0} \\ \mathbf{0} & \underline{Z}_2 & -\underline{Z}_3 \\ -\underline{Z}_1 & \mathbf{0} & \underline{Z}_3 \end{pmatrix}$$

$$\hat{\underline{\underline{Z}}}_p = \begin{pmatrix} \underline{Z}_{1p} & -\underline{Z}_{2p} & \mathbf{0} \\ \mathbf{0} & \underline{Z}_{2p} & -\underline{Z}_{3p} \\ -\underline{Z}_{1p} & \mathbf{0} & \underline{Z}_{3p} \end{pmatrix}$$

$$\hat{\underline{U}}_0 = \begin{pmatrix} \underline{U}_{120} \\ \underline{U}_{230} \\ \underline{U}_{310} \end{pmatrix}, \quad \hat{\underline{I}} = \begin{pmatrix} \underline{I}_1 \\ \underline{I}_2 \\ \underline{I}_3 \end{pmatrix}$$

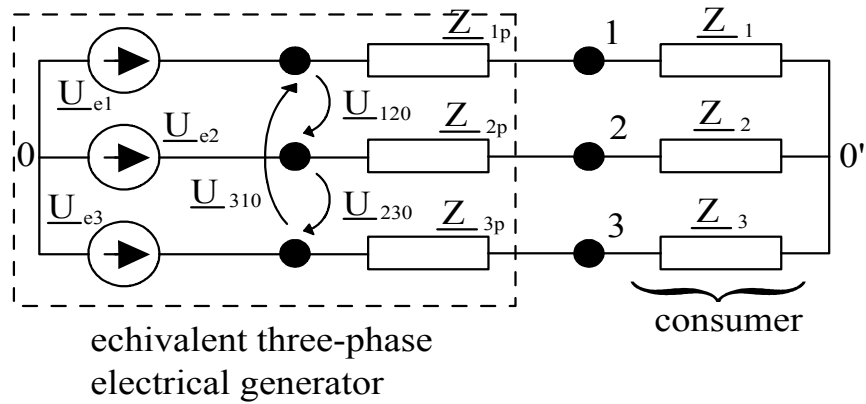

echivalent three-phase
electrical generator

consumer

*Figure 8. The electrical equivalent schema*

The matrix $\hat{\underline{\underline{Z}}}$ and $\hat{\underline{\underline{Z}}}_p$ are non-inverse. This property conduit to the impossibility to determinate the currents to consumer.

In order to determine the phase, fiction voltages $\underline{U}_{e1}^*, \underline{U}_{e2}^*, \underline{U}_{e3}^*$ we calculate the centre of gravity of the triangle formed by the potentials $\underline{V}_{10}, \underline{V}_{20}, \underline{V}_{30}$ of the phases 1, 2, 3, when let open, fig.9.a.

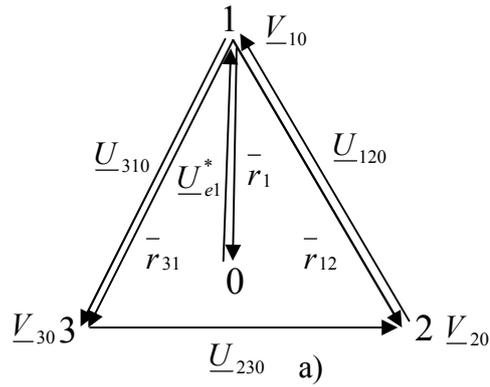

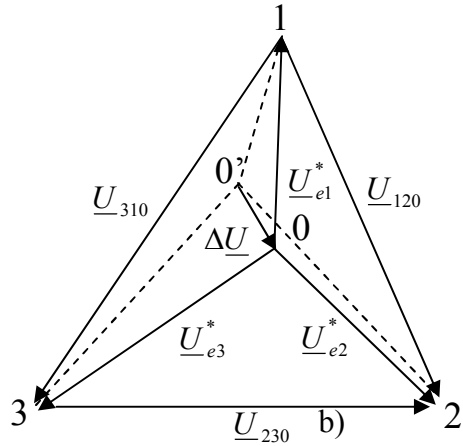

*Fig.9. Centre of gravity determinations for the line tensions triangle*

The position vector $\bar{r}_1$ of the centre of gravity O reported to the vertex 1 has the expression:

$$\bar{r}_1 = \frac{1}{3}\left(\bar{r}_{12} + \bar{r}_{31}\right) \tag{14}$$

where:

$$\begin{aligned} \bar{r}_1 &= -\underline{U}^*_{e1} \\ \bar{r}_{12} &= -\underline{U}_{120} \\ \bar{r}_{31} &= \underline{U}_{310} \end{aligned} \tag{15}$$

From (14) and (15) relations, we will obtain:

$$\underline{U}^*_{e1} = \frac{1}{3}\left(\underline{U}_{120} - \underline{U}_{310}\right) \tag{16}$$

By circular permutations, we can obtain the other phase, fiction phase's voltages:

$$\underline{U}^*_{e2} = \frac{1}{3}\left(\underline{U}_{230} - \underline{U}_{120}\right) \tag{17}$$

$$\underline{U}_{e3}^{*} = \frac{1}{3}\left(\underline{U}_{310} - \underline{U}_{230}\right) \qquad (18)$$

Neutral displacement of the consumer O' reported to the centre of gravity O, figure 9.b, is given by the relation:

$$\Delta\underline{U} = -\frac{\underline{U}_{e1}^{*}\underline{Y}_{1t} + \underline{U}_{e2}^{*}\underline{Y}_{2t} + \underline{U}_{e3}^{*}\underline{Y}_{3t}}{\underline{Y}_{1t} + \underline{Y}_{2t} + \underline{Y}_{3t}} \qquad (19)$$

where:

$$\begin{cases} \underline{Y}_{1t} = \dfrac{1}{\underline{Z}_{1p} + \underline{Z}_{1}} \\[2mm] \underline{Y}_{2t} = \dfrac{1}{\underline{Z}_{2p} + \underline{Z}_{2}} \\[2mm] \underline{Y}_{3t} = \dfrac{1}{\underline{Z}_{3p} + \underline{Z}_{3}} \\[2mm] \underline{Y}_{4t} = \dfrac{1}{\underline{Z}_{4p} + \underline{Z}_{4}} \end{cases} \qquad (20)$$

The matrix of the phase voltages at the consumer is:

$$\hat{\underline{\underline{U}}}_{1230} = \begin{pmatrix} \underline{U}_{e1}^{*} + \Delta\underline{U} & 0 & 0 \\ 0 & \underline{U}_{e2}^{*} + \Delta\underline{U} & 0 \\ 0 & 0 & \underline{U}_{e3}^{*} + \Delta\underline{U} \end{pmatrix} \qquad (21)$$

The matrix of the internal impedance three-phased electrical generator has the form:

$$\hat{\underline{\underline{Z}}}_{123p} = \begin{pmatrix} \underline{Z}_{1p} & 0 & 0 \\ 0 & \underline{Z}_{2p} & 0 \\ 0 & 0 & \underline{Z}_{3p} \end{pmatrix} \qquad (22)$$

And the one for the three-phased consumer:

$$\hat{\underline{\underline{Z}}}_{123} = \begin{pmatrix} \underline{Z}_{1} & 0 & 0 \\ 0 & \underline{Z}_{2} & 0 \\ 0 & 0 & \underline{Z}_{3} \end{pmatrix} \qquad (23)$$

The electrical three-phase currents we can write them as a diagonal matrix:

$$\hat{\underline{\underline{I}}}_{123} = \begin{pmatrix} \underline{I}_{1} & 0 & 0 \\ 0 & \underline{I}_{2} & 0 \\ 0 & 0 & \underline{I}_{3} \end{pmatrix} \qquad (24)$$

There is dependence between the electrical quantities that can be found in the Thèvenin theorem of the mono-phased, electrical circuits and the ones that can be found in the three-phased, electrical circuits without neutral, given by:

$$\underline{U}_{ab0} \Rightarrow \hat{\underline{\hat{U}}}_{1230} \tag{25}$$

$$\underline{Z}_{abp} \Rightarrow \hat{\underline{\hat{Z}}}_{123p}$$

$$\underline{Z} \Rightarrow \hat{\underline{\hat{Z}}}_{123} \tag{26}$$

$$\underline{I}_{abp} \Rightarrow \hat{\underline{\hat{I}}}_{123}$$

The Thèvenin theorem for the three-phased, electrical circuits without neutral has the expression:

$$\hat{\underline{\hat{U}}} = \left( \hat{\underline{\hat{Z}}}_{123p} + \hat{\underline{\hat{Z}}}_{123} \right) \hat{\underline{\hat{I}}}_{123} \tag{27}$$

The above relation generalize Thèvenin theorem of the mono-phased, electrical circuits:

$$\underline{U}_{ab0} = \left( \underline{Z}_p + \underline{Z} \right) \underline{I}$$

## 4. Three-phase system with neutral wire

A three-phase system with neutral wire can present in relation with a consumer from four terminal accesses:

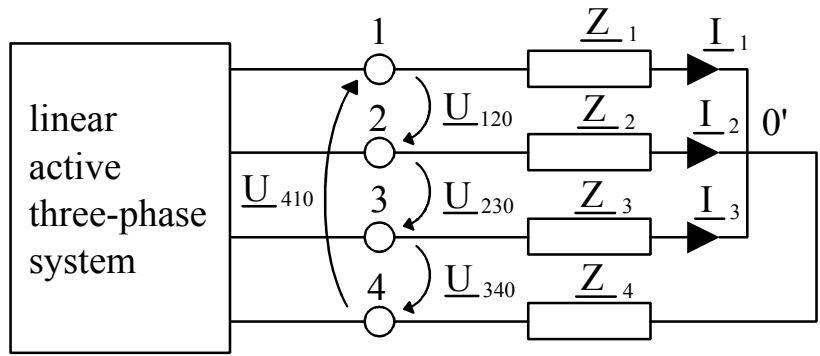

*Figure 10. Schema of the linear active three-phase system that is connect to a consumer*

These four access terminals 1, 2, 3 and 4' permit to assimilate the system with a multiphase system. The off tensions of the system in relation with the access terminals are $\underline{U}_{120}$, $\underline{U}_{230}$, $\underline{U}_{340}$ and $\underline{U}_{410}$.
In figure 11 is represented the equivalent electrical generator schema that corresponding to the four-access terminals system.

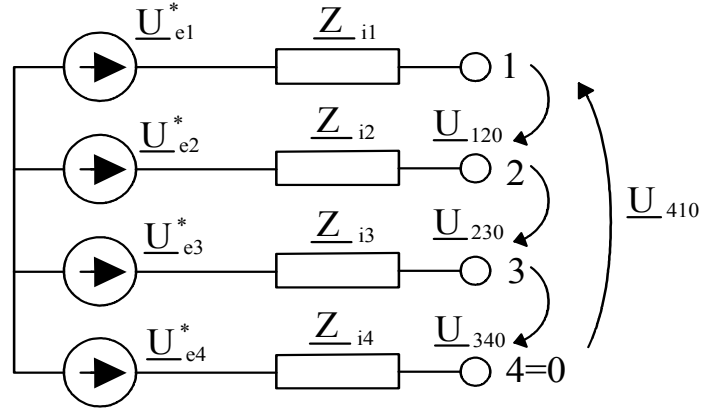

*Figure 11. The four-phase equivalent electrical generator schema*

In order to determine the phase fiction voltages $\underline{U}_{e1}^*, \underline{U}_{e2}^*, \underline{U}_{e3}^*, \underline{U}_{e4}^*$ we calculate the centre of gravity of the quadrilateral formed by the potentials $\underline{V}_{10}, \underline{V}_{20}, \underline{V}_{30}, \underline{V}_{40}$ of the phases 1, 2, 3, 4 when let open, figure 12, a, b.

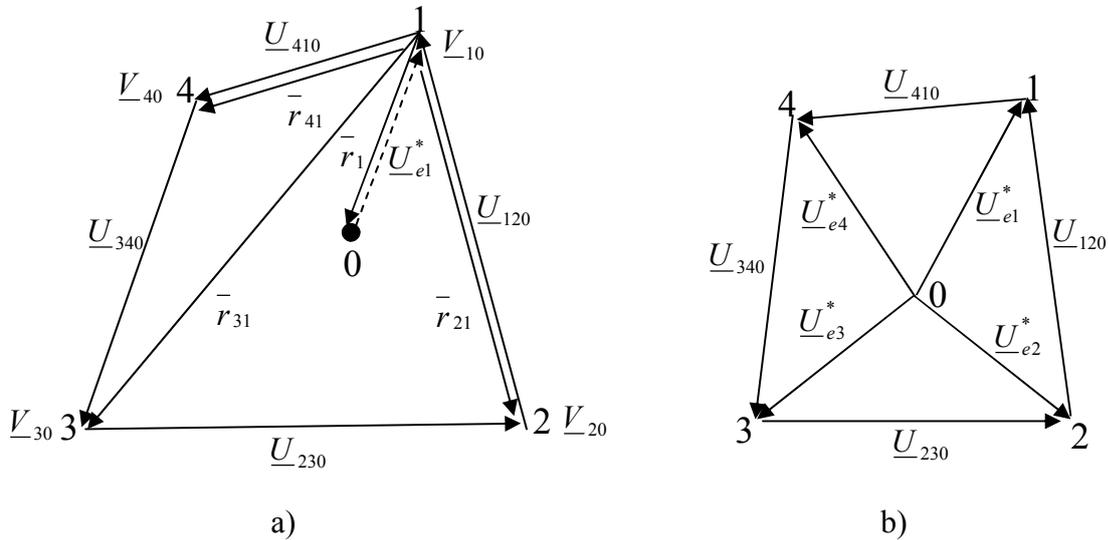

a)                 b)

*Figure 12. Centre of gravity determination of the quadrilateral*

The position vector $\bar{r}_1$ of the centre of gravity O, reported to the vertex 1 of the quadrilateral has the expression:

$$\bar{r}_1 = \frac{1}{4}\left(\bar{r}_{21} + \bar{r}_{31} + \bar{r}_{41}\right) \qquad (28)$$

Where:

$$\begin{cases} \bar{r}_1 = -\underline{U}_{e1}^* \\ \bar{r}_{21} = -\underline{U}_{120} \\ \bar{r}_{31} = \underline{U}_{410} + \underline{U}_{340} = -\underline{U}_{230} - \underline{U}_{120} \\ \bar{r}_{41} = \underline{U}_{410} \end{cases} \qquad (29)$$

From relations (28) and (29) we obtain (figure 12.a):

$$\underline{U}_{e1}^{*} = \frac{1}{8}\left(3\underline{U}_{120} + \underline{U}_{230} - \underline{U}_{340} - 3\underline{U}_{410}\right) \tag{30}$$

By circular permutations, we can obtain the other fiction phases-voltages:

$$\underline{U}_{e2}^{*} = \frac{1}{8}\left(3\underline{U}_{230} + \underline{U}_{340} - \underline{U}_{410} - 3\underline{U}_{120}\right)$$

$$\underline{U}_{e3}^{*} = \frac{1}{8}\left(3\underline{U}_{340} + \underline{U}_{410} - \underline{U}_{120} - 3\underline{U}_{230}\right) \tag{31}$$

$$\underline{U}_{e4}^{*} = \frac{1}{8}\left(3\underline{U}_{410} + \underline{U}_{120} - \underline{U}_{230} - 3\underline{U}_{340}\right)$$

The non-equilibrate consumer, with impedances $\underline{Z}_1, \underline{Z}_2, \underline{Z}_3, \underline{Z}_4$ connected in a star pattern generates the displacement of the neutral electrical potential of the consumer reported to the neutral electrical potential of the generator:

$$\Delta\underline{U} = -\frac{\underline{U}_{e1}^{*}\underline{Y}_{1t} + \underline{U}_{e2}^{*}\underline{Y}_{2t} + \underline{U}_{e3}^{*}\underline{Y}_{3t} + \underline{U}_{e4}^{*}\underline{Y}_{4t}}{\underline{Y}_{1t} + \underline{Y}_{2t} + \underline{Y}_{3t} + \underline{Y}_{4t}} \tag{32}$$

Analogously to the way we have proceeded in the paragraph 3 we obtain:

$$\underline{\hat{U}}_{12340} = \begin{pmatrix} \underline{U}_{e1}^{*} + \Delta\underline{U} & 0 & 0 & 0 \\ 0 & \underline{U}_{e2}^{*} + \Delta\underline{U} & 0 & 0 \\ 0 & 0 & \underline{U}_{e3}^{*} + \Delta\underline{U} & 0 \\ 0 & 0 & 0 & \underline{U}_{e4}^{*} + \Delta\underline{U} \end{pmatrix} \tag{33}$$

$$\underline{\hat{Z}}_{1234p} = \begin{pmatrix} \underline{Z}_{1p} & 0 & 0 & 0 \\ 0 & \underline{Z}_{2p} & 0 & 0 \\ 0 & 0 & \underline{Z}_{3p} & 0 \\ 0 & 0 & 0 & \underline{Z}_{4p} \end{pmatrix} \tag{34}$$

$$\underline{\hat{Z}}_{1234} = \begin{pmatrix} \underline{Z}_{1} & 0 & 0 & 0 \\ 0 & \underline{Z}_{2} & 0 & 0 \\ 0 & 0 & \underline{Z}_{3} & 0 \\ 0 & 0 & 0 & \underline{Z}_{4} \end{pmatrix} \tag{35}$$

$$\underline{\hat{I}}_{1234} = \begin{pmatrix} \underline{I}_{1} & 0 & 0 & 0 \\ 0 & \underline{I}_{2} & 0 & 0 \\ 0 & 0 & \underline{I}_{3} & 0 \\ 0 & 0 & 0 & \underline{I}_{4} \end{pmatrix} \tag{36}$$

In the above relations, $\overline{U}_{ei}, i = 1,2,3,4$, have the forms (30) and (31), $\Delta \underline{U}$ has the form (32), $Z_{ip}, i = 1,2,3,4$ have the forms (20).

For electrical three-phase circuits with neutral, the Thèvenin theorem has the form:

$$\hat{\underline{U}} = \left( \hat{\underline{Z}}_{1234p} + \hat{\underline{Z}}_{1234} \right) \hat{\underline{I}}_{1234} \qquad (37)$$

## 5. Conclusions

1. The Helmholtz – Thèvenin theorem for a linear active dipole may be generalized to a linear active three-phase system.
2. The obtained formula (27) and (37) generalized the Helmholtz –Thevenin' theorem.
3. The system currents determination its make by transformation $\Delta$ -Y.

**REFERENCES**

[1] Pl.Andronescu, *Electrotechnics Bases,* vol.II*,* Ed. Didactica si Pedagogica, Bucuresti, 1972.
[2] K.Simonyi, *Theoretical Electrotechnics,* Ed. Tehnica, Bucuresti, 1974.
[3] I.Antoniu, *Electrotechnics Bases,* vol.II*,* Ed. Didactica si Pedagogica, Bucuresti, 1974.
[4] A.Timotin, *Lessons Electrotehnics Bases,* Ed. Didactica si Pedagogica, Bucuresti, 1970.